\documentclass[final,3p,times,twocolumn]{elsarticle}

\usepackage{lineno,hyperref}
\modulolinenumbers[5]

\journal{Nuc. Instr. Meth. Nuc. Res. - Section B}

\begin{document}

\begin{frontmatter}

\title{MultiSIMNRA: a computational tool \\for self-consistent ion beam analysis using SIMNRA}

\author{T. F. Silva\fnref{myfootnote}}
\fntext[myfootnote]{Corresponding author. e-mail: tfsilva@if.usp.br}
\author{C.L. Rodrigues}
\author{M.V. Moro}
\author{G.F. Trindade}
\author{F.R. Aguirre}
\author{N. Added}
\author{M.A. Rizzutto}
\author{M.H. Tabacniks}
\address{Instituto de Física da Universidade de S\~ao Paulo, Rua do matão, trav. R 187,\\ 05508-090 S\~ao Paulo, Brazil.}
\author{M. Mayer}
\address{Max-Planck-Institut für Plasmaphysik, Boltzmannstr. 2, D-85748 Garching, Germany.}

\begin{abstract}
SIMNRA is widely adopted by the scientific community of ion beam analysis for interpretation of nuclear scattering analysis. Taking advantage of its recognized reliability and quality of the simulations, we developed a computer program that use parallel sessions of SIMNRA to perform self-consistent analysis for energy spectra of a given sample obtained using different techniques or experimental setups. In this paper, we present a result using MultiSIMNRA on self-consistent analysis for a multielemental thin film produced by magnetron sputtering. The results demonstrate the potentialities of the self-consistent analysis and its feasibility when using MultiSIMNRA.
\end{abstract}


\end{frontmatter}

%\linenumbers

\section{INTRODUCTION}

Computational tools always played an important role in the data interpretation of ion beam analysis (IBA) techniques during its historical development [1]. Since the early times many advances occurred in the comprehension of physical processes and measurement system effects, always followed by advances in the corresponding computational modeling [2].

Currently, SIMNRA is a widely adopted software by the IBA community [3]. Its strength lies on trusted modeling of the physical processes involved in the scattering calculation and measurements system effects. It was recently reported that upgrades can be expected for its new version concerning the skewness of energy spread distributions, improved handling of reaction cross-sections with structure, generalized layer roughness, and sample porosity [4].

During the last decade, one approach that has become a trend in the field and represents a great advance for the development of the analysis software is the simultaneous analysis of multiple measurements of ion beam techniques, the so called self-consistent analysis. This approach has gained attention since it ensures the reliable and unequivocal modeling of the sample [5], but equally important, the self-consistent analysis inherits the accuracy of the most accurate component of the analysis [6,7].

Taking advantage of the recognized reliability and quality of the simulations provided by SIMNRA, we developed a program for self-consistent analysis based on SIMNRA calculations. MultiSIMNRA uses computational algorithms to minimize an objective function running multiple instances of SIMNRA, finding the set of parameters that best fits simultaneously all experimental data. A more detailed information about MultiSIMNRA and its methods shall be published in the near future. In this paper, in order to demonstrate its potentialities and feasibility, we present a result obtained using MultiSIMNRA for the self-consistent analysis of a multielemental thin film produced by magnetron sputtering.

\section{METHODS}

\subsection{Sample description}

The sample consisted of a thin film of a mixture of aluminum, titanium and tantalum deposited on top of an amorphous carbon substrate using magnetron sputtering [ref]. Previous analysis found contamination of the elements hydrogen, carbon and oxygen in the film. The aim of the present analysis is to determine the contamination depth profiles.

\subsection{Experimental measurements}

We use different techniques to obtain the depth profile for each element and contaminants in this sample. For the quantification of the major elements in the film (Al, Ti and Ta), a usual Rutherford Back-scattering Spectrometry (RBS) analysis provided the quantification and depth profile. However, for the contaminants it was necessary to perform specific measurements to increase the sensitivity. Determination of O and C with their respective depth profiles were obtained using Elastic Back-scattering Spectrometry (EBS) with non-Rutherford resonant cross-section for different energies of incident beam. An Energy Recoil Detection Analysis (ERDA) analysis was used for the detection of the element H.

These measurements were divided into four irradiations with a $^4$He beam. Detectors recording the energy spectra at two different scattering angles were used to collect a total of eight spectra. In the first three irradiations, surface barrier detector detectors (both with 20-keV energy resolution) were placed at scattering angles of 170$^\circ$ (1.46-msr solid angle) and 120$^\circ$ (0.731-msr solid angle) wherein three different beam energies (2.200, 3.044 and 4.274-MeV) were used to probe the sample. The fourth irradiation (also at 2.2-MeV) was used in a ERDA configuration to enable the detection of recoiled H atoms. In this setup, the beam incidence angle on the sample was 80$^\circ$ and one detector was placed at a scattering angle of 20$^\circ$ with a 12-mm thick aluminum foil as absorber for stopping elastically scattered He (in this configuration the detector solid angle is 0.485 msr and the detector increases to 65-keV due to the angular straggling). Simultaneously to this FRS measurement, another detector was placed at 170$^\circ$ scattering angle to collect data of the RBS measurement at grazing angle increasing depth resolution, and completing the set of eight spectra.

\subsection{Basic physical data}

Two important sources of uncertainty in IBA techniques are the stopping power and the cross-sections data. In the special case of the self-consistent approach the accuracies of these databases are critical since inaccuracies inevitably lead to discrepancies in the joint interpretation of multiple spectra.

Thus, the choice of accurate databases for stopping and cross-section data is of major importance. For that, MultiSIMNRA has the option to use SRIM stopping powers for all spectra under analysis. Despite the good internal database of SIMNRA, the use of SRIM [10] data is preferable for traceability purposes, since its results are intensively confronted to experimental data and vice-versa [11]. For the cross-sections case, evaluated data provided by SigmaCalc [12] can be obtained in [13,14] and incorporated in the SIMNRA reaction list. After that, the appropriate cross-section must be configured in each of the SIMNRA files. The beam energy in the EBS analysis (4.274-MeV for C and 3.044-MeV for O) was adjusted to the position of the resonance in the cross-sections for each case, increasing the sensitivity for quantification (approximately 100 times greater for C and 20 times greater for O at the surface). SigmaCalc data were also used in the calculation of the FRS measurements since the p(a,p) a recoil cross-section deviates from Rutherford being approximately 10 times higher at 2.2 MeV. 

\subsection{MultiSIMNRA features}

In order to fit the set of experimental data, the MultiSIMNRA code minimizes the mean reduced $\chi^2$ as an objective function. This is evaluated as follows:

\begin{equation}
\chi^2_{red} = \frac{1}{S}\sum _{s=1}^{S} \left (   { \frac{1}{N_s-P_s-P_p} \cdot \sum_{channels} {\frac{(M_i-T_i)^2}{\sigma_i}}} \right ) 
\end{equation}
where $M_i$ and $T_i$ are respectively the calculated and experimental value of counts in channel $i$ of a spectrum $s$, $S$ is the total number of spectra, $N_s$ is the number of channels in the fitting region of the spectrum $s$, $P_s$ is the number of fitting parameters in the setup configuration for the measurement of the spectrum $s$, and $P_p$ is the number of fitting parameter in the sample depth profile.

The use of the reduced $\chi^2$ as objective function ensures that all spectra in the analysis have the same statistical weights in the end of the minimization process. In principle, in the beginning of the convergence it is expected that  spectra with more integrated counts have larger statistical weight but, it is also expected that, when the simulated data converges to the experimental data, this weights converges to the same values (all reduced $\chi^2$ values tends to the unity). What means that, this objective function gives priority for the spectra with larger integrated counts in the beginning of the minimization process but converges for the same statistical weights when getting closer to the minimum.

\begin{figure*}[!htb]
\centering
\includegraphics[width=16cm]{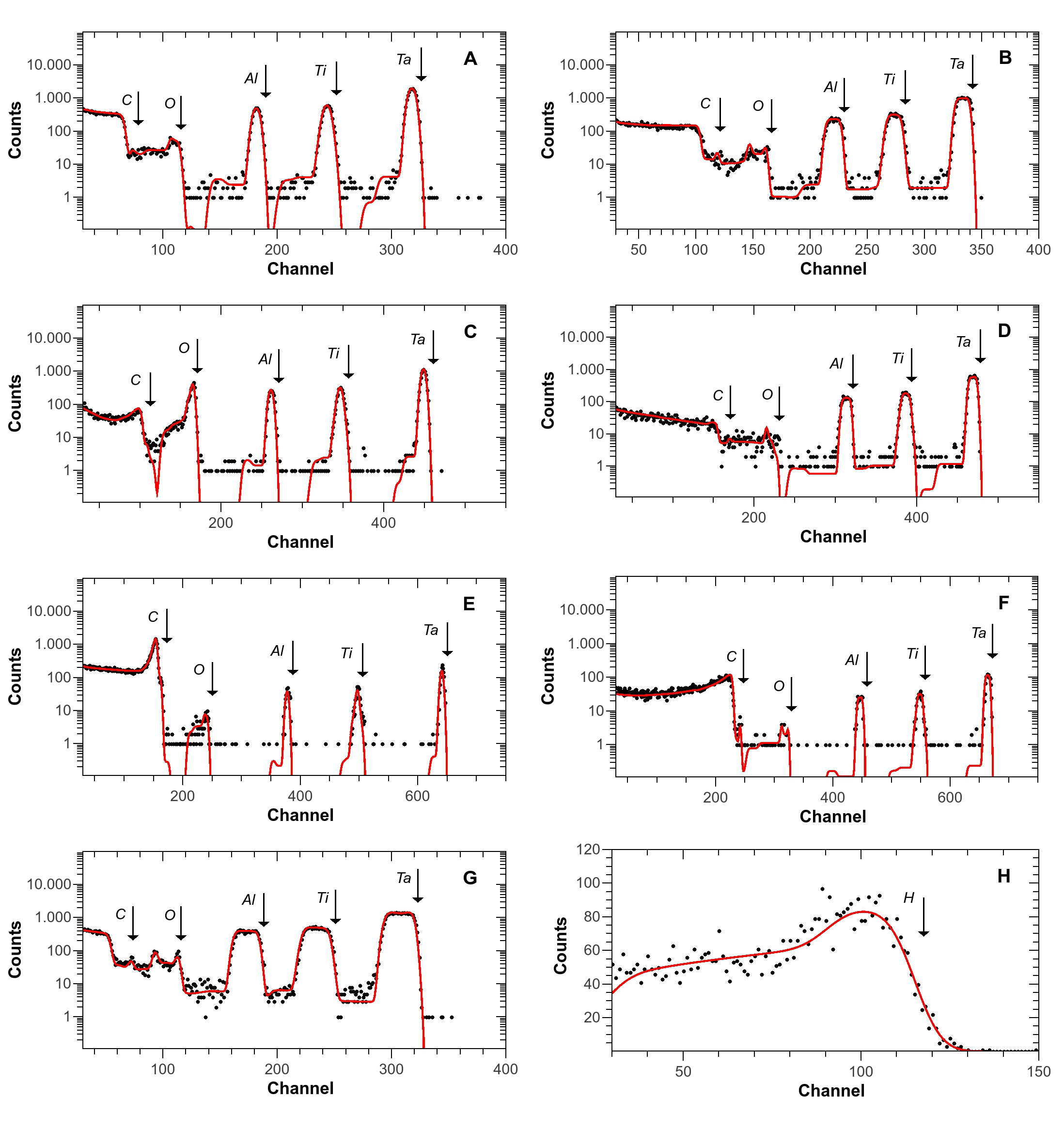}
\caption{A thin film of a mixture of Al, Ti and Ta analyzed using multiple techniques (the dots are experimental data) and the continuous line are the fit using MultiSIMNRA. A, C, E and G spectra were obtained with detector placed at 170$^\circ$ scattering angle. B, D and F spectra were obtained with detector at 120$^\circ$ scattering angle. And spectra G, the detector was placed at 20$^\circ$ scattering angle. Incidence angle for the G and H spectra was 80o. Beam energies: A and B, 2.2-MeV; C and D, 4.274-MeV; E and G, 3.044-MeV; G and H, 2.2-MeV.}
\label{fig_1}
\end{figure*}

MultiSIMNRA counts with four minimization algorithms to minimize the mean reduced $\chi^2$ of the set of spectra. In the case of this study, Simplex algorithm was used, and Evolutionary Annealing Simplex for refinements. Are also available: Differential Evolution and Levemberg-Marquart. The first and the latter are useful when the number of fitting parameters is low, and the other two have a larger convergence time, but are more efficient when the number of fitting parameters is high.

\begin{figure}[!htb]
\centering
\includegraphics[width=6cm]{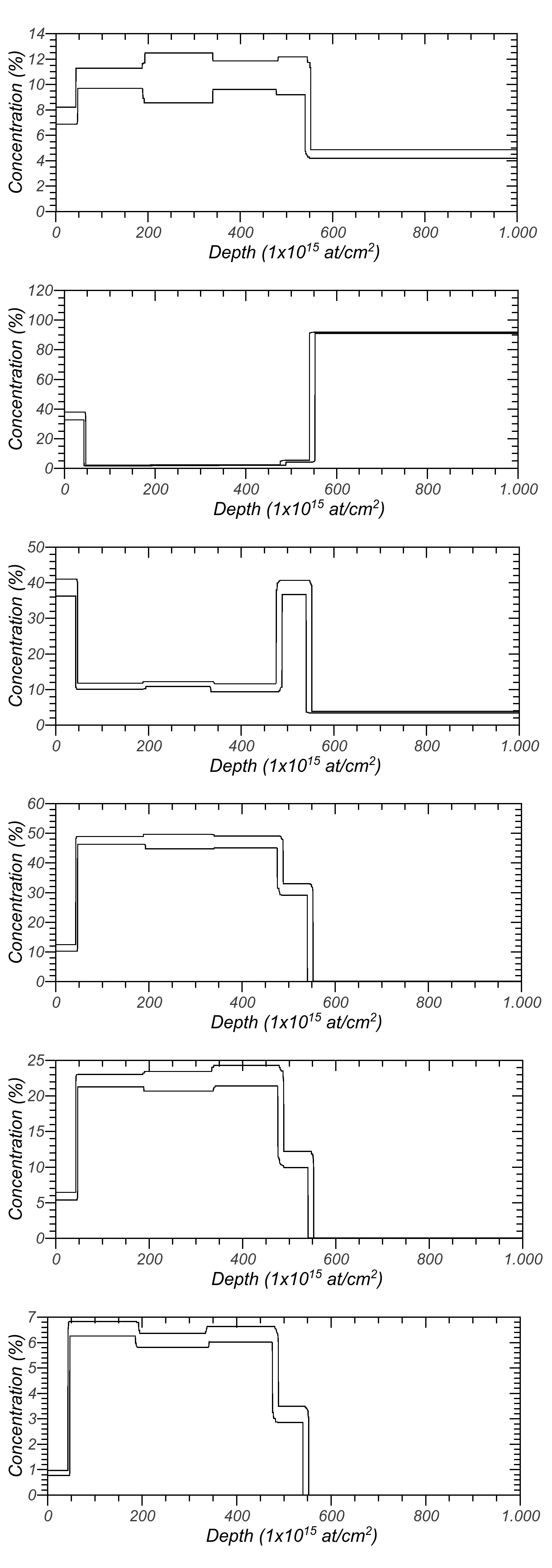}
\caption{Depth profile obtained using MultiSIMNRA that best fits all experimental data of Fig. 1. The two lines in each graph represent minimum and maximum limits defined by uncertainties calculated based on statistical significance of all spectra.}
\label{fig_2}
\end{figure}

The features of MultiSIMNRA were designed to provide to the user a smooth experience in self-consistent analysis using SIMNRA. One of these features is the definition of constraints by the definition of Links. A Link is a feature that says to MultiSIMNRA the relationship between fitting parameters. In the case of this analysis, the detector electronics calibration, gain and offset for the spectra collected by the same detector were linked, reducing the number of fitting parameters from 16 to 4. Also, for the EBS measurements the beam energy was left as fitting parameter within a range of 0.3\% of the energy read out. The beam energy of simultaneous measures was linked, reducing these number of fitting parameters from 8 to 4. The definition of Links in MultiSIMNRA is important not just because of the reduction of the convergence time by the direct reduction of fitting parameters, but also to establish conditions and reductions in degrees of freedom for the fitting process. The definition of links tying the fitting parameters increases the accuracy of the analysis. Another possibility of the Link feature is to correlate element concentrations simulating a chemical bound, but this feature was not used in the present analysis.

\section{RESULTS}

Fig. 1 shows the set of eight spectra with the corresponding SIMNRA calculations according to the MultiSIMNRA fitted parameters. The corresponding depth profiles for the contaminants are shown in Fig. 2. Besides, the beam energy for the EBS spectra was derived by the fit, and deviation of 0.24\% for the 3.044-MeV case and a 0.08\% for the 4.274-MeV case were found. These are excellent values and the fitting uncertainties lie under 0.05%.

%\begin{figure*}[!b]
%\centering
%\includegraphics[width=12cm]{Grafico1.jpg}
%\caption{A thin film of a mixture of Al, Ti and Ta analyzed using multiple techniques (the dots are experimental data) and the continuous line are the fit using MultiSIMNRA. A, C, E and G spectra were obtained with detector placed at 170$^\circ$ scattering angle. B, D and F spectra were obtained with detector at 120$^\circ$ scattering angle. And spectra G, the detector was placed at 20$^\circ$ scattering angle. Incidence angle for the G and H spectra was 80o. Beam energies: A and B, 2.2-MeV; C and D, 4.274-MeV; E and G, 3.044-MeV; G and H, 2.2-MeV.}
%\label{fig_1}
%\end{figure*}

\section{DISCUSSIONS}

The RBS measurements (spectra A and B) can give good results for Al, Ti and Ta quantification, but have low statistics for the determination of the precise concentrations of C and O. These elements can be assessed using EBS measurements taking advantages of the resonances that occur at 4.400 MeV and 3.038-MeV beam energies for scattering cross-sections of He in C (spectra C and D) and O (spectra E and F) respectively. Additionally, an RBS measurement at grazing angle (spectrum G) can provide good depth resolution for profile determination for all elements and the ERDA measurement (spectrum H) complements the analysis providing the concentration of H. Handling the data simultaneously and self-consistently, it was possible to obtain the depth profile for contaminants presented in Fig.2. It is possible to observe the oxidation of the film in the interfaces, and the diffusion of H through the substrate interface (which lies approximately at 500 TFU).

%\begin{figure}[!htb]
%\centering
%\includegraphics[width=7cm]{Grafico2.jpg}
%\caption{Depth profile obtained using MultiSIMNRA that best fits all experimental data of Fig. 1. The two lines in each graph represent minimum and maximum limits defined by uncertainties calculated based on statistical significance of all spectra.}
%\label{fig_2}
%\end{figure}

\section{CONCLUSIONS}

MultiSIMNRA can be a powerful tool to fit multiple energy spectra of scattering techniques using SIMNRA calculations. The easy-to-use-philosophy implemented in MultiSIMNRA and inspired by SIMNRA can provide a smooth experience for the user. Besides, the Link feature provides a possibility to include prior knowledge into the analysis.

\section{ACKNOWLEDGMENTS}
The authors would like to thank Brazilian Funding Agencies – FAPESP (Process number: 2012/00202-0), CNPq and CAPES – for their support to this research.

%\bibliography{mybibfile}

\section*{References}

\begin{thebibliography}{99}
%1
\bibitem{Ikeoka}
R.A. Ikeoka, C.R. Appoloni, M.A. Rizzutto, T.F. Silva and A.M. Bandeira, PIXE analysis of pre-colonial pottery from Sambaqui do Panaquatira, 13th PIXE conference, (2013).

%2
\bibitem{Ambiel}
V. C. Ambiel, M.A. Rizzutto, J.F. Curado, P.H.O.V. Campos, E.A.M. Kajiya and T.F. Silva, Metallic objects belonging to the first emperor of Brazil studied with PIXE technique, 13th PIXE conference, (2013).

%3
\bibitem{Moro}
M.V. Moro, T.F. Silva, N. Added, G.F. Trindade, M.H. Tabacniks, Study of Boron detection limit using in-air PIGE setup at LAMFI-USP, AIP Conference Proceedings, 1625, 120-124 (2014).

%4
\bibitem{Rizzutto}
M.A. Rizzutto, M.V. Moro, T.F. Silva, G.F. Trindade, N. Added, M.H. Tabacniks, E.M. Kajiya, P.H.V. Campos, A.G. Magalhães, M. Barbosa, External-PIXE analysis for the study of pigments from a painting from the Museum of Contemporary Art,  Nucl. Instr. Meth. B, 332, 2014, 411-414.

%5
\bibitem{Santos}
H.C. Santos, N. Added, T.F. Silva, C.L. Rodrigues, External-RBS, PIXE and NRA analysis for ancient swords, Nucl. Instr. Meth. B, 345, 15, (2015), 42-47.


\end{thebibliography}

\end{document}